\documentclass[showpacs,twocolumn,floatfix,prb]{revtex4}
\usepackage{graphicx} 	
\usepackage{bm} 	
\usepackage{amssymb}
\usepackage{amsmath}
\usepackage{amsfonts}
\usepackage{epstopdf}
\usepackage{times}

\begin{document}

\title{Strongly anharmonic current-phase relation in ballistic graphene Josephson junctions}

\author{Annica M. Black-Schaffer}
 \affiliation{NORDITA, Roslagstullsbacken 23, SE-106 91 Stockholm, Sweden}
\author{Jacob Linder}
\affiliation{Department of Physics, Norwegian University of Science and Technology, N-7491 Trondheim, Norway}

\date{Received \today}
\begin{abstract}
Motivated by a recent experiment directly measuring the current-phase relation (CPR) in graphene under the influence of a superconducting proximity effect, we here study the temperature dependence of the CPR in ballistic graphene SNS Josephson junctions within the the self-consistent tight-binding Bogoliubov-de Gennes (BdG) formalism. By comparing these results with the standard Dirac-BdG method, where rigid boundary conditions are assumed at the S$\mid$N interfaces, we show on a crucial importance of both proximity effect and depairing by current for the CPR. 
The proximity effect grows with temperature and reduces the skewness of the CPR towards the harmonic result. In short junctions ($L<\xi$) current depairing is also important and gives rise to a critical phase $\phi_c<\pi/2$ over a wide range of temperatures and doping levels.
\end{abstract}


\maketitle

\section{Introduction}
Ever since the discovery of graphene\cite{Novoselov04} there has been interest in merging the unique two dimensional massless relativistic Dirac fermion spectrum of graphene with superconductivity. A few years ago, graphene superconductor-normal metal-superconductor (SNS) Josephson junctions were successfully fabricated.\cite{Heersche07, Miao07,Shailos07, Du08} Here superconductivity is induced in the S regions of the graphene through proximity to superconducting contacts deposited on top of these regions, and, through the Josephson effect, a finite supercurrent can exist even in the N region of the graphene. 
One of the key properties of a SNS junction is the current-phase relation (CPR), and, very recently, the first direct measurement of the CPR in graphene SNS junctions appeared.\cite{Chialvo10} 

Theoretically, ballistic graphene SNS junctions were studied even before the first experimental results. Several novel features were discovered, such as specular Andreev reflection,\cite{Beenakker06} and a finite supercurrent in undoped graphene at zero temperature, despite the point-like Fermi surface.\cite{Titov06} These results, along with most theoretical results on the Josephson effect in graphene SNS junctions,\cite{Moghaddam06, Cuevas06, Maiti07,Titov07} have employed the Dirac-Bogoliubov-de Gennes (DBdG) formalism, where the standard BdG formulation is applied to the Dirac spectrum.\cite{Beenakker06} To solve the resulting equations, rigid boundary conditions are assumed for the superconducting order parameter $\Delta$, such that $\Delta$ takes on a fixed, non-zero value in S, whereas it is zero in N. However, such an approach explicitly ignores any processes in which $\Delta$ is reduced in the S regions, such as proximity effect or depairing by current (see e.g.~Ref.~[\onlinecite{Golubov04}] for a review). Moreover, with the Josephson current in the DBdG approach usually calculated as the current carried by sub-gap Andreev bound states, this method is limited to junction lengths $L<\xi$, where $\xi$ is the superconducting coherence length. 
Other, related methods have been employed to relax some of the constraints in the original works, but they still apply rigid boundary conditions for $\Delta$.\cite{Gonzalez08,Burset08,Hagymasi10}
In fact, in order to not apply any boundary conditions on $\Delta$ at the S$\mid$N interface, a self-consistent treatment is needed. One such method is the self-consistent tight-binding (TB) BdG formalism, where one only assumes that the superconducting contacts induces a pairing potential into the graphene and then solve self-consistently for $\Delta$ in the whole SNS structure.\cite{Nikolic01,Covaci06,Black-Schaffer08} Not only does this approach give an explicit calculation of the full proximity effect, including current depairing, but it also results in a Josephson current appropriately calculated from this proximity effect. Previous works by the authors\cite{Black-Schaffer08, Linder09}  employing this formalism at zero temperature showed on some corrections to the CPR compared to the DBdG results, as well as deviations for the critical current as function of junction length.

Now, with experimental data on the CPR in graphene at hand, it is of large interest to theoretically map out the temperature dependence of the CPR in graphene. Very recently, some theoretical temperature dependent CPR results appeared using the DBdG approach,\cite{Hagymasi10} but otherwise all theoretical work have been at zero temperature.
The goals of this work are thus twofold: (1): Establish the correct CPR in ballistic graphene SNS Josephson junctions as function of temperature using the self-consistent TB BdG formalism. (2): Determine how well the DBdG approach captures the CPR, and, if applicable, determine the source(s) of discrepancy.

We will here show that both the proximity effect depletion of superconductivity in the S regions and depairing by current are large in short junctions ($L<\xi$). In fact, these effects lead to a CPR where the critical current is reached for a phase $\phi_c<\pi/2$. This is true over a large temperature range and even more prominent for high doping levels in the graphene. In order to capture these effects a self-consistent solution of $\Delta$ is crucial since rigid boundary conditions explicitly ignores any such processes in the S regions.
For longer junctions ($L>\xi$) the proximity effect depletion is unchanged but depairing by current is now not a big issue, and here $\phi_c\geq \pi/2$. In this junction length regime the DBdG approach using Andreev bound states is not formally justified, and due to a large proximity effect, a self-consistent approach is still needed.

The rest of the article is organized as follows. In the next section we briefly introduce the methods used, both the self-consistent approach and the DBdG equation using rigid boundary conditions. Then in Section III we report our results, including the temperature dependent CPR for different junctions as well as an analysis of the influence of proximity effect and current depairing on the CPR. Finally, we summarize our key findings and comment on the applicability of our results to current and future experiments.

%
\section{Method}
Our starting point is the nearest neighbor tight-binding Hamiltonian on the graphene lattice together with an on-site $s$-wave superconducting order parameter $\Delta_i$, which is induced by the proximity to external superconducting contacts:
%
\begin{align}
\label{eq:H}
\mathcal{H} = &-t\sum_{\langle i,j\rangle,\sigma} (a_{i \sigma}^\dagger b_{j \sigma}  + \textrm{H.c.}) +\sum_{i \sigma} \mu_{i \sigma}(a_{i \sigma}^\dagger a_{i \sigma} + b_{i \sigma}^\dagger b_{i \sigma})\notag\\
&+ \sum_i \Delta_i (a_{i \uparrow}^\dagger a_{i \downarrow}^\dagger + b_{i \uparrow}^\dagger b_{i \downarrow}^\dagger).
\end{align}
Here $a_{i\sigma}^\dagger$ ($b_{i \sigma}^\dagger$) creates an electron with spin $\sigma$ on the A (B) sublattice in unit cell $i$. $t \sim 2.5$~eV is the nearest neighbor hopping amplitude, whereas $\mu_i$ is the chemical potential. We will assume that, due to the external contacts, the chemical potential is rather large and constant in the S regions whereas it can be tuned (to a constant value) with a back gate voltage in the N region.

\subsection{DBdG equation}
For an analytical treatment within the DBdG framework we set $\Delta_N = 0$, whereas $\Delta_S = \Delta(T)$ with $\Delta(T = 0) = \Delta_0$. Here $\Delta(T)$ is set to the standard BCS temperature dependence. This amounts to applying rigid boundary conditions on $\Delta_i$ at the S$\mid$N interface. By treating the S and N regions separately, the above Hamiltonian can be Fourier transformed and the kinetic energy linearized around the Dirac points to produce the standard DBdG equation:\cite{Beenakker06}
%
\begin{align}
\label{eq:DBdG}
&\begin{pmatrix}
H_0-\mu & \Delta \\
\Delta^\dagger & -H_0 +\mu \\
\end{pmatrix} \Psi = \epsilon \Psi,
\end{align}
where $\Delta$ and $\mu$ take on different, but constant, values in the S and N regions, as described above.
Moreover, $H_0 = -i \hbar v_F (\sigma_x \partial_x + \sigma_y \partial_y)$, where $\hbar v_F = \sqrt{3}ta/2$ with $a = 2.46$~\AA~being the lattice constant of graphene. 

The strategy for calculating the Josephson current in the junction is to first obtain the energy spectrum for the Andreev-bound states in the N region. This is done by matching the wave functions given by Eq.~(\ref{eq:DBdG}) at the two S$\mid$N interfaces and then solve for the allowed energy eigenstates, $\varepsilon$.
The general expression for the Josephson current can then be written as:\cite{Beenakker91b}
\begin{align}\label{eq:jos}
I(\phi) = \frac{4e}{\hbar} \sum_{\pm}\sum_n \frac{\partial \varepsilon_{\pm,n}}{\partial \phi} f(\varepsilon_{\pm,n}),
\end{align}
where the summation over $\pm$ denotes all Andreev levels and the summation over $n$ denotes the transverse momentum index. Here, $\phi$ is the phase drop across the junction,$f(x)$ is the Fermi-Dirac distribution function, and the prefactor of 4 is due to the spin- and valley-degeneracy.
%
Let us first consider the case of $\mu_S = \mu_N$, i.e. no Fermi level mismatch (FLM) in the junction. In the wide junction limit where the junction width $W$ satisfies $W\gg L$, we may replace the summation over discrete transverse momentum indices with an integral as follows:
\begin{align}
\sum_n \to \int \frac{\text{d}k_y}{2\pi/W} = \frac{\mu_N W}{2\pi\hbar v_F} \int \text{d}\theta\cos\theta,
\end{align}
where the integration over the angle $\theta$ takes into account all possible trajectories.
The Andreev levels in a SNS junction with no FLM have the form:\cite{Beenakker91b}
\begin{align}
\varepsilon_\pm &= \pm \Delta(T)\sqrt{1 -\tau\sin^2(\phi/2)}
\end{align}
where for graphene $\tau$ has the specific form:
\begin{align}
\tau &= \frac{\cos^2\theta}{1 - \sin^2\theta\cos^2[\mu_N L\cos\theta/(\hbar v_F)]}.
\end{align}
Inserting the above equations in to Eq.~(\ref{eq:jos}) and introducing the normalization constant $I_0 = (W/\xi)e\Delta_0/\hbar$, where $\xi=\hbar v_F/\Delta_0$ is the superconducting coherence length, we arrive at the expression:
\begin{align}
\frac{I(\phi)}{I_0} &= \frac{\mu_N}{2\pi\Delta_0} \int^{\pi/2}_{-\pi/2}  \text{d}\theta \frac{[\Delta(T)]^2}{\Delta_0\varepsilon_+} \frac{\tau\sin\phi\cos\theta}{\tanh^{-1}(\beta\varepsilon_+/2)},
\end{align}
where $\beta$ is the inverse temperature.
Finally turning our attention to the FLM case, a similar procedure\cite{Titov06} leads to the expression:
\begin{align}
\frac{I(\phi)}{I_0} &= \frac{\Delta(T)\xi}{\Delta_0W} \sum_n^\infty \frac{\tau_n\sin\phi}{\sqrt{1-\tau_n\sin^2(\phi/2)}},
\end{align}
where $\tau_n$ is given by Eq.~(17) in Ref.~[\onlinecite{Titov06}].

\subsection{Self-consistent treatment}
In the self-consistent numerical treatment we do not predetermine the value of $\Delta_i$ nor apply rigid boundary conditions at the S$\mid$N interface. Instead we assume that the influence of the superconducting contacts on the underlying graphene is only through an induced attractive pairing potential. For $s$-wave pairing the simplest potential is a constant, non-zero, attractive Hubbard-$U$ term in the S regions. By applying the mean-field approximation to the attractive Hubbard model we reproduce Eq.~(\ref{eq:H}) but with an added self-consistency condition
%
\begin{align}
\label{eq:selfcons}
\Delta_i = -U_i \frac{\langle a_{i\downarrow} a_{i\uparrow}\rangle + \langle b_{i\downarrow} b_{i\uparrow}\rangle}{2},
\end{align}
where $U_i = U$ in S but zero in N.
We can now solve self-consistently for $\Delta_i$ by first guessing a profile for $\Delta_i$ throughout the SNS junction, solving Eq.~(\ref{eq:H}) with this guess, recalculating $\Delta_i$ using Eq.~(\ref{eq:selfcons}), and then reiterate this process until two subsequent $\Delta_i$ profiles are within a predetermined error margin. 
In order to study the proximity effect between the S and N regions in the graphene we focus on the pairing amplitude $F_i = \Delta_i/U_i$. While $\Delta_i$ will only be non-zero in the S regions, $F_i$ can, due to proximity effect leakage from S to N, be non-zero even in N. Alongside this leakage also comes a depletion of $F$ in S such that $F < \Delta_0/U$ close to the interface. It is this latter process that, as we will show below, significantly changes the CPR. 
The other important property is the Josephson current $I$ that flows through the junction if there exists a phase difference $\phi$ in $\Delta_i$ across N. From the continuity equation for the charge current we can, using the self-consistent solution for $\Delta_i$, calculate $I$. For additional details on this self-consistent method see Ref.~[\onlinecite{Black-Schaffer08}].

\subsection{Simulation details}
We will assume clean, smooth interfaces such that a Fourier transform along the direction parallel to the interface is applicable. In the self-consistent treatment, the type of interface can be varied but with an on-site pairing the direction of the interface will not matter. We will here use the zigzag interface, such that one unit cell is $\sqrt{3}a/2$ long. Moreover, we will only consider the wide junction regime, $W\gg L \sim \xi$. In the self-consistent treatment the junctions are naturally infinitely wide, due to Fourier transforming along the interface, whereas in the DBdG approach $W = 30\xi$ is used. For a direct comparison between the two methods the current will be given in units of $W/\xi$.

In this article we will focus on a few representative values for the various physical input parameters. 
In the superconducting S regions we use $U = 1.36t$ and $\mu_S = 0.6t$. 
These values give $\Delta_0 = 0.042t$, and a superconducting coherence length $\xi = \hbar v_F/\Delta_0 \approx 24$~unit cells. This satisfies $\xi \ll \lambda_F = \hbar v_F/\mu$ in both S and N, a requirement for the DBdG solution, and will also allow us to self-consistently investigate both the $L<\xi$ and $L>\xi$ regimes, where $L$ is the length of the N region. Moreover, with these values we get S regions as small as 50 unit cells displaying clear bulk behavior, which is advantageous due to the the computational demands of the self-consistent method. Note though that for a direct comparison with an experimental setup $\Delta_0$ is very large. However, we believe that our results are applicable even for smaller $\Delta_0$,\cite{Black-Schaffer08} and we, therefore, explicitly only report our results in dimensionless units.
We will vary the chemical potential in the N region from the Dirac point ($\mu_N = 0$) to moderately doped ($\mu_N = 0.47\mu_S$), to no FLM at the interface ($\mu_N = \mu_S$).
In addition, we will investigate both short junctions with $L = 0.42\xi$, where our DBdG approach is technically justified, and long junctions with $L = 2.1\xi$.

Finally, let us comment on the application of a phase gradient in $\Delta_i$. We apply a phase difference $\Delta \phi$ between the superconducting order parameters in the outermost regions of the two S regions. In the self-consistent treatment the phase is, however, allowed to relax also in the S regions close to the interfaces. Thus the phase drop $\phi$ across the junction itself, i.e.~across N, will necessarily satisfy $\phi\leq \Delta\phi \leq \pi$ in the self-consistent treatment, with an inequality in the first step as soon as the supercurrent is non-zero. As a consequence, we will not be able to trace out the self-consistent CPR for $\phi$ close to $\pi$ in junctions with high currents. While this appears as a numerical artifact in this context, it is, in fact, closely related to the physical $2\pi$ phase-slip process in Josephson junctions (see, e.g., Ref.~[\onlinecite{Tinkhambook}]).
%
\section{Results} 
We will start by reporting the CPR for several different ballistic graphene SNS junctions. Then we will explicitly analyze the critical current and phase, followed by a qualitative analysis of the effect of proximity effect and current depairing on the CPR. The latter two quantities are captured by our self-consistent approach but is not included within the DBdG framework.

\subsection{CPR}
\begin{figure}[htb]
\includegraphics[scale = 0.96]{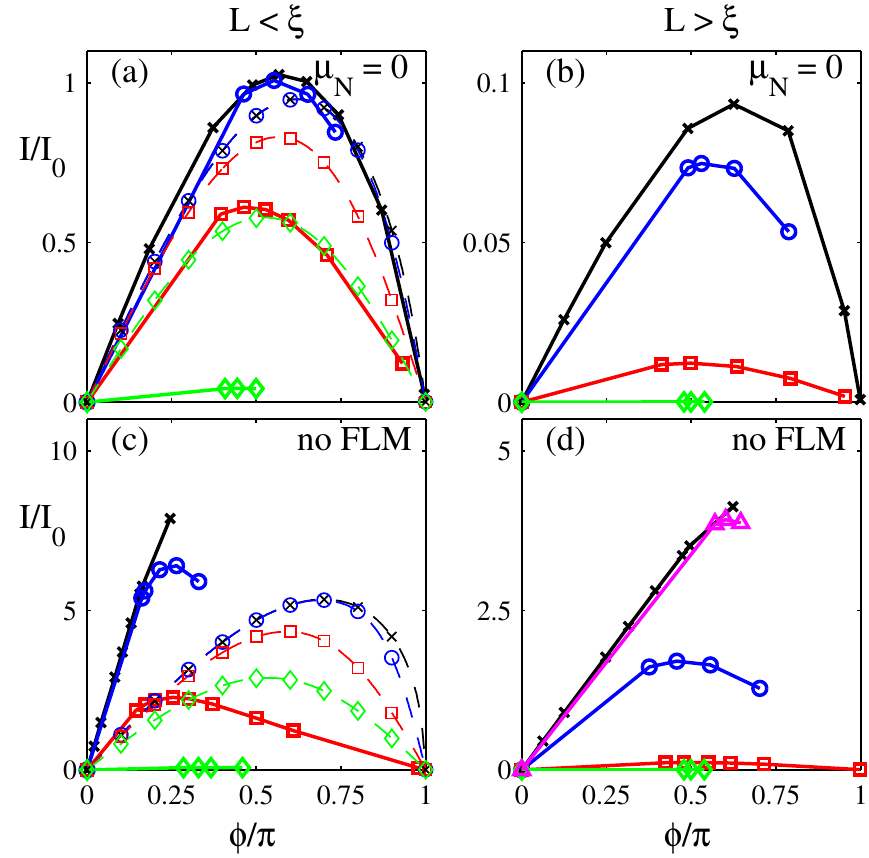}
\caption{\label{fig:Ivsphi} (Color online) Josephson current $I$ in units of $I_0 = (W/\xi)e\Delta_0 /\hbar$, when $v_F = 1$, as a function of the phase drop $\phi$ across the junction for $\mu_N = 0$ (a, b) and no FLM (c, d) for short junctions (a, c) and long junctions (b, d) at $T/T_c = 0$ (black, $\times$), 0.14 (blue, $\circ$), 0.43 (red, $\square$), and 0.87 (green $\diamond$). In (d) $T/T_c = 0.025$ (magenta, $\triangle$) is also plotted whereas it falls on top of the $T = 0$ results in the other plots. Self-consistent TB BdG results (solid, lines are only guides to the eye), and DBdG results (dashed).
}
\end{figure}

\begin{figure}[htb]
\includegraphics[scale = 0.96]{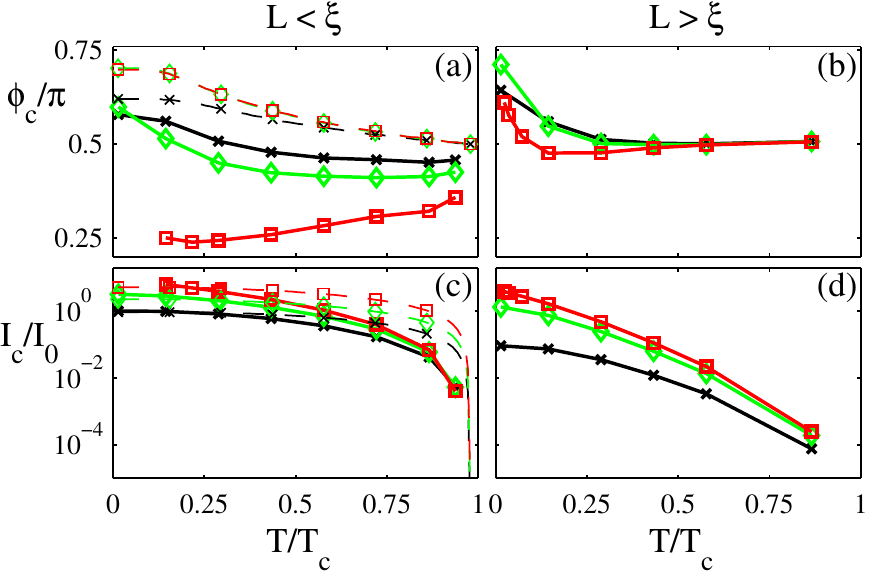}
\caption{\label{fig:Icphic} (Color online) Critical phase $\phi_c/\pi$ (a, b) and critical current $I_c/I_0$ (c, d) as a function of the reduced temperature $T/T_c$ for $\mu_N/\mu_S = 0$ (black, $\times$), 0.47 (green, $\diamond$), and no FLM (red, $\square$) for short junctions (a, c) and long junctions (b, d). Self-consistent results (solid, lines are only guides to the eye), and DBdG results (dashed). Note the log scale in (c, d).
}
\end{figure}

Figure \ref{fig:Ivsphi} shows the CPR for four representative cases with $\mu_N = 0$ and no FLM, and for short and long junctions, respectively. For each case we plot the CPR at four different temperatures, $T/T_c = 0, 0.14, 0.43,$ and 0.87.
For short junctions we also report the DBdG results as dashed curves for a straightforward comparison.
%
%
We will comment more specifically on the skewness, or anharmonicity, of the CPR when extracting $\phi_c$ in Fig.~\ref{fig:Icphic}, but we see directly that the DBdG results only approximately reproduces the self-consistent results in the limit of low temperature and low doping levels in N. For increasing doping levels and/or increasing temperatures, the self-consistent results have a skewness towards $\phi_c<\pi/2$, whereas $\phi_c \geq \pi/2$ is always the case for the DBdG results. By parameterizing the skewness as $S = 2\phi_c/\pi-1$, we will refer to the former case as negative skewness.
We also directly see that the total current in the junction is naturally increased with increasing doping level, decreasing junction length, or decreasing temperature. Note though that the self-consistent results show a much larger suppression of the current with increasing temperature than the DBdG results.

Before we continue, the case of no FLM at the interface deserves special attention. Here there is no interface barrier and we thus have an SNS junction with fully transparent interfaces. Such junctions, independent on junction length, have been shown to have a $1/T$ diverging superconducting decay length $\xi_N$ in the normal region.\cite{Covaci06} For $T = 0$ this is directly manifested in a linear drop of $\Delta \phi$ throughout the whole junction (i.e. even in the S regions) and, consequently, also a linear CPR, resulting in $\phi_c \rightarrow \pi$.\cite{Black-Schaffer08} The diverging $\xi_N$ at $T = 0$ also leads to currents not limited by tunneling through Andreev bound states but by the size of the (induced) superconducting gap in the N region. Due to these high currents we cannot fully trace out the CPR in these junctions at low temperatures. However, in the accessible range, the self-consistent CPR is indeed linear, which supports the claim that $\phi_c \rightarrow \pi$ as $T\rightarrow 0$. Note that this numerical problem can partially be circumvented by studying longer junctions, as seen in Fig.~\ref{fig:Ivsphi}(d), since these junctions carry smaller currents due to a smaller superconducting gap in the N region.
Since the current calculation in the DBdG framework explicitly assumes the presence of a finite number of Andreev bound states in the normal (metal) N region, this method cannot capture the effects of the diverging $\xi_N$. This results in an underestimation of both the skewness and the absolute value of the critical current in junctions with no FLM at very low temperatures when using the DBdG method. 

Let us now focus on the critical phase $\phi_c$, extracted for three different doping levels in Figs.~\ref{fig:Icphic} (a, b).
%
In short junctions the DBdG results (dashed lines) show a notable (positive) skewness away from the traditional harmonic Josephson relation $I= I_c \sin(\phi)$. At the Dirac point at zero temperature, $\phi_c = 0.63\pi$ as already established analytically,\cite{Titov06} but with increasing temperature the CPR approaches the harmonic form, in agreement with other recent DBdG results.\cite{Hagymasi10}
 The skewness in the DBdG approach is somewhat increased with increasing doping but the overall temperature dependence is not highly sensitive to the doping level in N, nor is the no FLM case distinctly different from the other results.
 What is most striking with Fig.~\ref{fig:Icphic}(a) is the fact that the DBdG results do {\it not} reproduce the self-consistent results reliably.
 At low doping levels, $\phi_c$ is essentially uniformly shifted to lower values in the self-consistent approach, such that $\phi_c<\pi/2$ above some temperature. The temperature at which $\phi_c = \pi/2$ decreases with increasing doping levels. This leads to a substantial negative skewness with increasing doping, and thus the self-consistent results deviate even more from the DBdG results as the doping increases in the N region. At very high temperatures there is finally a tendency to approach a harmonic CPR, which is the development for any ballistic SNS junction with rigid boundary conditions as $T \rightarrow T_c$.\cite{Golubov04} Within our numerical accuracy we cannot, however, model temperatures close enough to $T_c$ to formally see if $\phi_c \rightarrow \pi/2$ as $T\rightarrow T_c$.
 
For long junctions the skewness is instead in general positive or only minutely negative for the highest doping levels. Also, $\phi_c \rightarrow \pi/2$ even for only moderately high temperatures. 
We can for $L> \xi$ not formally justify the use of our DBdG framework, but a comparison still shows on a large discrepancy between the self-consistent results and the DBdG results, although it is smaller than for shorter junctions. Notably, the self-consistent results reach the harmonic $\phi_c = \pi/2$ result much faster than the DBdG results. 

 In terms of the critical current, even with the log-scale in Figs.~\ref{fig:Icphic}(c, d), we see that a self-consistent treatment results in a significantly lower current for all junctions except at very low doping levels and temperatures. 
In fact, the discrepancy between the two methods grow strongly with increasing temperature. 
  
 In aggregate, the above results directly leads to two conclusions: (1) a self-consistent approach is crucial in the short junction regime and still important for longer junctions. (2) a no FLM junction at zero temperature has a diverging superconducting decay length $\xi_N$ and is therefore fully superconducting. This leads to a linear CPR with $\phi_c \rightarrow \pi$ as $T \rightarrow 0$.
The large difference between the self-consistent and the DBdG results must stem from processes that are not captured when applying rigid boundary conditions to $\Delta_i$ at the S$\mid$N interface. We have already discussed the special case of no FLM junctions at zero temperature where $\Delta_N$ becomes finite. Also, very generally, processes influencing the superconducting state in the S region are present in all junctions and at all temperatures.
These processes include proximity effect depletion of the pairing amplitude $F$ in the S region close to the interface and the additional loss of pairs in the whole junction because of a finite supercurrent, or, equivalently, a finite phase gradient. 
In fact, shifts of $\phi_c$ to the region $\phi_c\geq \pi/2$ are mainly governed by processes in the N region, as evident in the DBdG approach, whereas shifts towards $\phi_c\leq \pi/2$ occur because of processes in the S regions.\cite{Golubov04} That is, in the absence of the loss of Cooper pairs in the S regions of the interface, the self-consistent results would reproduce the DBdG results, where the intrinsic properties of the graphene N region lead to $\phi_c \geq \pi/2$.

\subsection{Proximity effect}

\begin{figure}[htb]
\includegraphics[scale = 0.95]{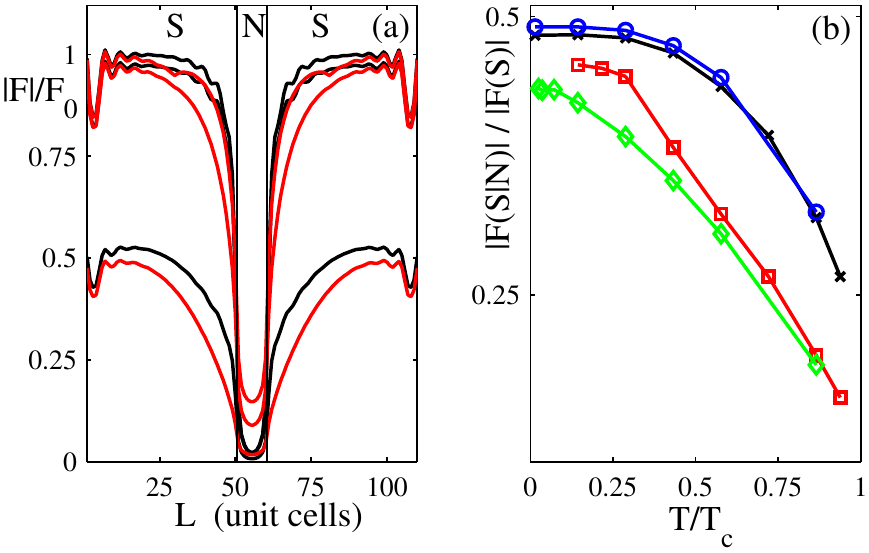}
\caption{\label{fig:proxeff} (Color online) (a): Absolute value of the pairing amplitude $F$ in units of $F_0 = \Delta_0/U$ at the critical current in a short junction for $\mu_N = 0$ (black) and no FLM (red) for temperatures $T/T_c =  0, 0.43$, and 0.87 (decreasing amplitude). Vertical lines mark the S$\mid$N interfaces. (The curve for no FLM at T = 0 is at the highest achievable $\phi$, which did not reach $I_c$.)
(b): Ratio of $|F|$ at the S$\mid$N interface to $|F|$ in the bulk as a function of the reduced temperature $T/T_c$ for $\mu_N$ = 0 in a short junction (black, $\times$) and long junction (blue, $\circ$) and for no FLM in a short junction (red, $\square$) and long junction (green, $\diamond$).
}
\end{figure}

Figure \ref{fig:proxeff} contains the evidence for proximity effect depletion of pair amplitude at the interfaces.
%
In (a) we see that for strong doping (red) the proximity effect is clearly larger, with more depletion of $F$ in S and the accompanied accumulation of pairs in N. This is expected since with increasing doping in N, the effective barrier at the interface due to the FLM is decreased. The increase in the Josephson current with doping of N is a direct consequence of this enhancement of $F$ in N. 
In terms of temperature dependence, the pair amplitude reduction in the bulk S regions with increasing temperature follow the standard BCS temperature dependence and is thus also included in the DBdG treatment. 
On top of this, there is, however, an additional loss of pair amplitude both on the S and N sides of the S$\mid$N interfaces with increasing temperature. This latter process is only captured in the self-consistent method. 
Also seen in (a) are oscillations in $F_i$ at both ends of the S regions. These are due to the outer edges of the S regions, but do not influence the results of the junction itself. The small oscillations for $\mu_N = 0$ at the interfaces at low temperatures are directly correlated with Friedel-like charge oscillations, which are only present when N is close to the Dirac point.\cite{Black-Schaffer08}
 
 In (b) we analyze the proximity effect at the interface more closely. For strong doping in N (red, green), the proximity effect is notably stronger than for an undoped N region (black, blue). However, note that the proximity effect is essentially independent on the junction length $L$. This directly tells us that the stronger discrepancy between the DBdG and the self-consistent results found for short junctions is not due to the proximity effect.
Moreover, there is also a significantly increased interface proximity effect with increasing temperature. This explains the stronger decrease in both the critical current and phase with increasing temperatures than found in the DBdG results.

\subsection{Current depairing}
\begin{figure}[htb]
\includegraphics[scale = 0.95]{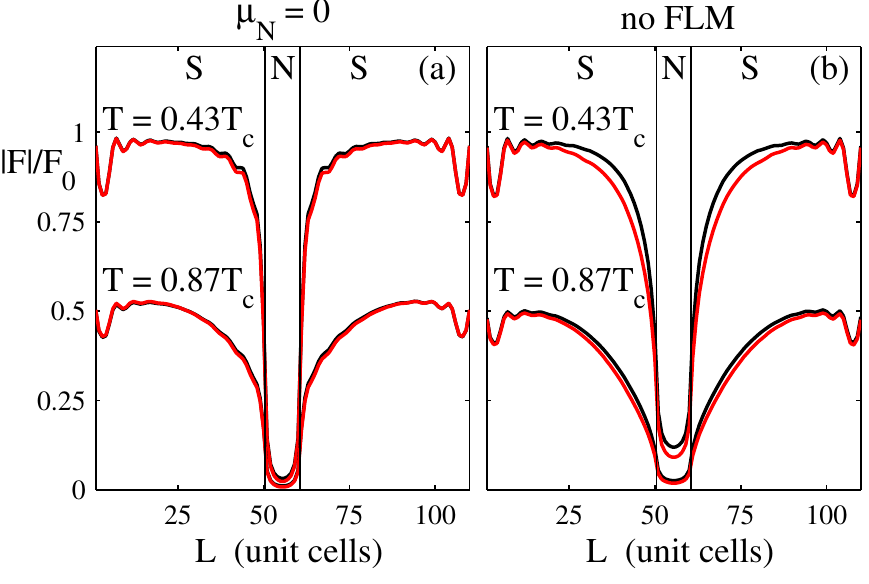}
\caption{\label{fig:currdepair} (Color online) Absolute values of the pairing amplitude $F$  in units of $F_0 = \Delta_0/U$ in short junctions for $\phi = 0$ (black) and $\phi = \phi_c$ (red) at two different temperatures for $\mu_N = 0$ (a) and no FLM (b). Vertical lines mark the S$\mid$N interfaces.
}
\end{figure}

Finally in Fig.~\ref{fig:currdepair} we study the absolute value of the pair amplitude $|F|$ for both $\phi = 0$ and $\phi_c$ in short junctions.  A difference in $|F|$ for these two phase gradients will directly signal the influence of a finite phase gradient or, by extension, a supercurrent on $F$. In the case of a loss of $|F|$ going from $\phi = 0$ and $\phi_c$ this is known as depairing by current, a process present in any junction carrying a finite supercurrent. 
%

In (a) we see that for low doping levels in N, there is a small amount of current depairing at low temperatures, which decreases with increasing temperatures. For heavily doped N regions (b), the depairing is significantly larger, though it also decreases with increasing temperature.
The same set of curves for long junctions show {\it no} change with applied phase gradient and thus depairing by current is only an issue in short junctions. This directly leads to the conclusion that it is current depairing which is responsible for the smaller and even negative skewness found in shorter junctions. Its doping dependence also explains why the skewness is more negative for larger doping levels. Moreover, the strong increase in proximity effect with increasing temperature is partially counteracted in short junctions by the decreased current depairing. Thus we see less of a temperature dependence for the skewness when the junction length is decreased.
In passing, we also note that the strength of the depairing process continues to increase as $\phi$ is increased from $\phi_c$ towards $\pi$, despite a decrease in the supercurrent. Thus the depairing is technically a function of the finite phase difference over the junction and not per see of the current, although they are intimately related.

%
\section{Conclusions} 

We have above self-consistently established the temperature dependent CPR in ballistic graphene SNS Josephson junctions with varying doping levels and junction lengths. We have also compared our results with the currently prevalent method of studying these junctions, the so-called DBdG method, which explicitly relies on the use of rigid boundary conditions for the superconducting order parameter $\Delta$ at the S$\mid$N interfaces.
The DBdG method gives a critical phase $\phi_c = 0.6-0.7$ at zero temperature, with larger values for higher doping levels in N, which slowly then reduces to the harmonic value of $\phi_c = \pi/2$ as $T \rightarrow T_c$. Our DBdG results are formally only valid in the short junction regime ($L< \xi$), but other recent work within the same DBdG framework has shown on a similar behavior for long junctions.\cite{Hagymasi10}
The self-consistent results, however, are in most junctions qualitatively different from the DBdG results. First of all, any process influencing the superconducting state in the S regions of the junction can only be captured when the rigid boundary conditions are removed and $\Delta$ is calculated self-consistently. Such processes are known\cite{Golubov04} to decrease the skewness of the CPR and can even produce negative skewness, i.e.~give $\phi_c < \pi/2$.
We have studied both the influence of proximity effect depletion of $\Delta$ in S close to the interfaces and the additional loss of Cooper pairs when a finite phase gradient is applied across the junction. The latter process, is known as depairing by current and is present whenever there is a finite supercurrent in the junction. 
The proximity effect is important and modifies the CPR in all junctions we have studied. It increases with doping and, even more so, with temperature. It is, however, independent of the junction length. Current depairing, on the other hand, is only important in short junctions. It increases with doping, but decreases with temperature, both natural consequences of the amount of current in the junction. In fact, current depairing helps explain the growing discrepancy in $I_c$ with decreasing $L$ between the analytical and self-consistent results found in Ref.~[\onlinecite{Linder09}].
Based on these results we can qualitatively explain the modified CPR in the self-consistent results: 
For long junctions only the proximity effect modifies the DBdG results. The proximity effect reduces the skewness significantly and also heavily suppresses the critical current, especially at higher temperatures. However, for $T$ approaching $T_c$, $\phi_c$ is pinned to the harmonic result, thus implying that in the very high temperature limit rigid boundary conditions are still applicable.
In shorter junctions, the same proximity effect is at play, but here current depairing causes further decrease in $\phi_c$. This is especially true for higher doping levels where we find $\phi_c$ as small as $\pi/4$. The temperature dependence for the current depairing is opposite that of the proximity effect, and we thus see a somewhat smaller temperature variation in $\phi_c$ for shorter junctions. Also, for short junctions a noticeable anharmonicity is present closer to $T_c$, which by itself strongly signals that these junctions are not well modelled using rigid boundary conditions.\cite{Golubov04}
In addition to these processes, junctions with no FLM at the interfaces have $\phi_c \rightarrow \pi$ as $T \rightarrow 0$, due to a $1/T$ diverging superconducting decay length $\xi_N$.\cite{Covaci06,Black-Schaffer08} Since the DBdG framework assumes $\Delta_N = 0$, also this effect is only captured in our self-consistent treatment.

Let us finally also briefly comment on the applicability of our results to an experimental setup. 
In the recent experiment measuring the CPR in graphene SNS junctions,\cite{Chialvo10} the skewness was found to always be positive and increasing linearly with the critical current (i.e.~with decreasing temperature and/or increasing doping). It was, however, concluded that the junctions were very likely in the quasi-diffusive regime and, thus, an explicit comparison with our work is not possible. 
Nonetheless, in Ref.~\onlinecite{Hagymasi10} a qualitative comparison was made to ballistic DBdG results in the appropriate long junction regime ($L \sim 3.5 \xi$), which also showed linear behavior for small $I_c$, although the slope did not match the experimental results.
Our results show that, even in this long junction limit, self-consistency is necessary in ballistic junctions in order to correctly capture the CPR, due to the proximity effect. The temperatures used in the experiment was, however, large enough that within our current model setup we cannot study these high temperatures, since $\Delta(T)$ becomes too small. Still, we can see that at slightly lower temperatures the skewness is not hugely temperature or doping dependent. We thus speculate that when relaxing the ballistic requirement, proximity effect corrections become less important. 
Ballistic transport has, however, already been demonstrated in graphene SNS junctions with the measurement of multiple Andreev reflections.\cite{Miao07} Thus, a measurement of the CPR in ballistic graphene junctions does not seem to be a too distant an achievement.
Our main, and easily verifiable, prediction in ballistic junctions -- that the skewness becomes negative due to a combination of current depairing and proximity effect in short junctions -- will therefore be important as the experimental junction lengths shrink in the future.

In summary, we have found that proximity effect, depairing by current, and diverging superconducting decay length all qualitatively modifies the CPR in graphene Josephson junctions. Thus, a correct description of the CPR necessarily requires a self-consistent treatment of the superconducting order parameter.
This is different from metallic SNS junctions where it has been shown that rigid boundary conditions are applicable for junctions with large FLM at the interface.\cite{Kupriyanov81, Golubov04}

%
\begin{acknowledgments}
A.M.B-S. thanks Nadya Mason for useful correspondence and Sebastian Doniach for providing the computational resources. 
\end{acknowledgments}

\bibliographystyle{apsrev}

\end{document}